\documentclass[11pt]{article}
\usepackage[numbers]{natbib}
\usepackage{times}
\usepackage{latexsym}
\usepackage{subfigure}
\usepackage{tikz}
\usepackage{forest}
\usepackage{tikz-dependency}
\usepackage[a4paper]{geometry}
\forestset{
dg edges/.style={for tree={parent anchor=south, child anchor=north,align=center,base=bottom,where n children=0{tier=word,edge=dotted}{}}},
}

\usetikzlibrary{positioning,arrows,decorations.pathreplacing,shapes}

\usepackage{setspace,caption}
\usepackage{authblk}

\usepackage{setspace}
\usepackage{times}
\usepackage{enumerate}

\usepackage{amssymb}
\usepackage{mathtools}
\usepackage{amsmath}
\usepackage{float}
\usepackage{lipsum}
\usepackage{graphicx,subfigure}
\usepackage{epstopdf}

\DeclareMathOperator*{\argmax}{argmax}
\DeclareMathOperator*{\argmin}{argmin}

\title{A decentralized route to the origins of scaling in human language}

\author[1]{Felipe Urbina\thanks{furbina@gmail.com}}
\author[2]{Javier Vera\thanks{jxvera@gmail.com.}}
\affil[1]{Facultad de Ciencias, Universidad Mayor, Camino La Pir\'amide, 5750,
Huechuraba, Santiago, Chile}
\affil[2]{Facultad de Ingenier\'ia y Ciencias, Universidad Adolfo Ib\'a\~{n}ez, Avda. Diagonal Las Torres 2640, Pe\~{n}alol\'en, Santiago, Chile}

\begin{document}
\maketitle



\begin{abstract}
The Zipf's law establishes that if the words of a (large) text are ordered by decreasing frequency, the frequency versus the rank decreases as a power law with exponent close to $-1$. Previous work has stressed that this pattern arises from a conflict of interests of the participants of communication. The challenge here is to define a computational multi-agent language game, mainly based on a parameter that measures the relative participant's interests. Numerical simulations suggest that at critical values of the parameter a human-like vocabulary, exhibiting scaling properties, seems to appear. The appearance of an intermediate distribution of frequencies at some critical values of the parameter suggests that on a population of artificial agents the emergence of scaling partly arises as a self-organized process only from local interactions between agents.
 \end{abstract}



 

\section{Introduction}

Human vocabularies are constrained by a fundamental statistical principle of organization, the so-called \textit{Zipf's law}, which establishes that the frequency of a word decays inversely as a power law of its rank \cite{Altmann2016,1742-5468-2017-1-014002,DBLP:books/degruyter/KAP2005,1367-2630-16-11-113010,Zipf:36,zipf1949human}. More precisely, if the words are ordered by decreasing frequency, the frequency of the $k$-th word, $P(k)$, follows

\begin{equation}
P(k)\sim k^{-\alpha}.
\end{equation}

This scaling law in the distribution of word frequencies has been understood in terms of a dichotomy between high frequency words, that require little memory effort (like the word ``the''), and low frequency words, that minimize the disambiguation effort of highly specialized concepts (like the word ``computer''). Zipf referred to the lexical trade-off between two competing pressures, \textit{ambiguity} and \textit{memory}, as the \textit{least effort principle}.

A family of recent works have stressed the understanding of Zipfian properties of language within an Information Theoretic analysis \cite{Cancho788,FerreriCancho2005,FerreriCancho2005b,CANCHO2006242,1742-5468-2007-06-P06009}. Remarkably, a crucial initial study of Ferrer-i-Cancho and Sol\'e in 2003 \cite{Cancho788} showed that Zipf's law is the outcome of the arrangement of word-meaning associations satisfying the simultaneous interests of both speakers and listeners. Their model results strongly suggested the appearance of a phase transition at a critical low effort stage in both competing pressures. Their study, however, only focused on the emergence of scaling as a simple optimization process operating on a single matrix of word-meaning associations, without any consideration about population structure or communicative interactions between speakers and hearers. 

Here, the work is guided thus by the following question (based on a related proposal of \cite{DBLP:journals/complexity/SoleCVS10}): Can artificial populations of agents develop, without any kind of central control, a shared vocabulary satisfying Zipfian properties? The main aim here is therefore to define a multi-agent language game \cite{1742-5468-2011-04-P04006}, according to different levels of word \textit{ambiguity} and \textit{memory} usage. Moreover, the hypothesis is that at some intermediate level of the participant's interests agents will share a word-meaning mapping exhibiting Zipfian scaling properties. The focus of this paper is an abstract \textit{decentralized} solution in which agents collectively reach shared communication systems without any kind of central control influencing the formation of a human-like language, and only from local conversations between few participants \cite{Steels95,Steels96,baronchelli_naming_jstat,steels2011REVIEW}.

The model proposed here is based on a prototypical agent-based model for computational studies of language formation, the \textit{naming game} \cite{Steels95,steels2011REVIEW,baronchelli_naming_jstat,1742-5468-2011-04-P04006}, which considers a finite population of agents, each one endowed with a memory to store, in principle, an unlimited number of words. At each discrete time step, a pair of agents, one speaker and one hearer, negotiate words as hypothesis for naming one object. Under the typical dynamics of the \textit{naming game}, the population will share after a finite amount of time a unique word for referring to the object. 


\section{Model}


\subsection{Main elements}

The game is played by a finite population of agents $P=\{1,...,p\}$, sharing both a set of words $W=\{1,...,n\}$ and a set of meanings $M=\{1,...,m\}$. A word-meaning mapping can be expressed by a $n\times m$ \textit{lexical matrix} $L=(l_{ij})$, where $l_{ij}=1$ if the $i$-th word conveys the $j$-th meaning, and $l_{ij}=0$, otherwise. In a related framework, lexical matrices are understood in terms of \textit{language networks} \cite{DBLP:journals/complexity/SoleCVS10}, in which there are two disjoint sets of \textit{nodes}, $W$ and $M$, and the \textit{edges} establish word-meaning relationships. 

The agent $k\in P$ is endowed in turn with its own word-meaning mapping, expressed by the $n\times m$ \textit{lexical matrix} $L^k=(l_{ij}^k)$, which is known by the agent $k$, and unknown by any other agent $k' \in P \setminus \{k\}$. 

Next, two technical concepts are introduced for the agent $k \in P$. The agent $k$ \textit{knows} the word $i \in W$ if $\sum_{j=1}^m l^k_{ij}\geqslant 1$, that is, the sum of the $i$-the row over the columns of the lexical matrix $L^k$ is greater or equal than 1. The quantity $a^k(i)=\sum_{j=1}^m l^k_{ij}$ is called the \textit{ambiguity} of the word $i$ for the agent $k$. 

Given a word $i \in W$, the value of its ambiguity $a^k(i)$ is closely related to the pressures competing in the Zipfian lexical trade-off. In the case of $a^k(i)=1$, indeed, the speaker's \textit{memory} is maximized, since the word $i$ is available only for one meaning $j^* \in M$, whereas the hearer's \textit{disambiguity} effort is minimized, since it faces the least effort of disambiguate the word-meaning association. At the opposite case, $a^k(i)=m$, the speaker's \textit{memory} is minimized, based on the fact that the word $i$ is associated to every meaning $j \in M$. The hearer's \textit{disambiguation} effort, on the contrary, is minimized, since the effort to establish a word-meaning association is maximized. 


Some \textit{repair} strategies of the agent's interactions need the introduction of the following notation: the $j$-th column of the lexical matrix $L^k$ is denoted $L^k_{\bullet j}$. With this notation, the \textit{Hadamard} product between $L^k_{\bullet j}$ and the canonical vector $e_i$ of dimension $n$ is defined as the entrywise product $L^k_{\bullet j} \odot e_i$. For example, the product between the vectors of dimension $n=5$, $L^k_{\bullet j} = (0,1,0,1,1)^T$ and $e_2=(0,1,0,0,0)^T$, is $L^k_{\bullet j} \odot e_i =(0,1,0,0,0)^T$.

\subsection{Basic interaction rule}

The dynamics of the language game is defined by pairwise speaker-hearer interactions at each discrete time step $t \geqslant 0$. At $t=0$, each agent $k \in P$ is endowed with a lexical matrix $L^k=(l^k_{ij})$, in which each entry $l^k_{ij}$ is equal to 1 or 0 with probability $0.5$. The basic interaction rule is defined by three steps at each time step $t$,

\begin{enumerate}[(\textbf{step} 1)]
\item a pair of agents is selected uniformly at random: one plays the role of \textit{speaker} $s$ and the other plays the role of \textit{hearer} $h$;

\item the speaker chooses uniformly at random one column (meaning) $j^* \in \{1,...,m\}$ from its lexical matrix $L^s$. Next, the speaker inspects its own lexical matrix $L^s$ in order to select a word associated to $j^*$, denoted $i^*$. There are two possibilities in this inspection:
 
\subitem (i) \textbf{if} $\sum_{i=1}^n l^s_{ij^*}=0$, the speaker chooses the word $i^*$ uniformly at random from $W$, and sets $l^s_{ij^*}=1$ in its lexical matrix $L^s$;

\subitem (ii) \textbf{otherwise}, the speaker \textit{calculates} the word $i^*$ based on its own lexical interests (that is, based on the of conflict between \textit{ambiguity} and \textit{memory} amount).

The speaker transmits the word $i^*$ to the hearer. 

\item the hearer behaves as in the \textit{naming game}. On the one hand, mutual agreement between the speaker and the hearer involves \textit{alignment} strategies \cite{steels2011REVIEW}. On the other hand, if the hearer does not know the word $i^*$, that is, $l^h_{i^*j^*}=0$, a \textit{repair} strategy is established in order to increase the chance of future agreements. More precisely,\\

\subitem (i) \textbf{if} $l^h_{i^*j^*}\neq 0$, both speaker and hearer updates the $j^*$-th column of their lexical matrices, by the \textit{Hadamard} products $(\odot)$.

\begin{center}
\begin{align*}
	L^s_{\bullet j^*} \leftarrow L^s_{\bullet j^*} \odot e_{i^*} \\
    L^h_{\bullet j^*} \leftarrow L^h_{\bullet j^*} \odot e_{i^*}
\end{align*}
\end{center}
    
\subitem (ii) \textbf{otherwise}, the hearer establishes a simple repair strategy: it adds 1 to the entry $(i^*j^*)$ of its lexical matrix $L^h$.

\end{enumerate}


\subsection{Relative interests of speakers and hearers}

What would be the minimal adaptation of the basic interaction rule that enables to include at the same time the interests of both speakers and hearers? In order to define relative interests, one feasible solution involves that speakers would prefer to transmit words associated to a relative ambiguity, defined by a simple relationship between the two extreme values of ambiguity ($a^k(i)=0$ or 1). The solution consists in the following version of the \textbf{step} 2:

\begin{enumerate}[(\textbf{step} 1)]\addtocounter{enumi}{1}

\item (i) \textbf{if} $\sum_{i=1}^n l^s_{ij^*}=0$, the speaker chooses the word $i^*$ uniformly at random from $W$, and sets $l^s_{ij^*}=1$ in its lexical matrix $L^s$;\\

\item[] (ii) \textbf{otherwise}, the speaker calculates $i^*$ according to the \textit{ambiguity parameter} $\wp \in [0,1]$. Let $random \in [0,1]$ be a random number. Then, \\
		\begin{itemize}
		\item \textbf{if} $random\geqslant\wp$, the speaker calculates $i^*$ as the least ambiguous word 
		\begin{equation*}
		i^*=\min_{\{w: l^s_{wj^*}\neq 0\}} \sum_{i=1}^n l^s_{wj^*};
		\end{equation*}\\
		\item \textbf{otherwise}, the speaker calculates $i^*$ as the most ambiguous word 
		\begin{equation*}
		i^*=\max_{\{w: l^s_{wj^*}\neq 0\}} \sum_{i=1}^n l^s_{wj^*}.
		\end{equation*}
		
		\end{itemize}		  

\end{enumerate}

\subsubsection*{Examples}

At some time step, consider the scenario in which the topic of the interaction is the meaning (column) $j^*=2$; and the speaker $k\in P$ is endowed with the lexical matrix 

$$L^k = 
\begin{pmatrix}
0 & 0 & 0 & 1 \\
0 & 1 & 1 & 0 \\ 
1 & 1 & 1 & 1 \\
1 & 0 & 1 & 1 \\   
\end{pmatrix}$$

Therefore, the speaker calculates one of the following words:

\begin{itemize}
\item the most ambiguous word (row) $i^{*}$ as $$i^*=\argmax_{\{i: l_{ij^{*}}^{k }\neq 0\}} \sum_{j=1}^m l_{ij}^k=\argmax_{i \in \{2,3\} } \sum_{j=1}^m l_{ij}^k=3$$
\item the least ambiguous word (row) $i^{*}$ as $$i^*=\argmin_{\{i: l_{ij^{*}}^{k }\neq 0\}} \sum_{j=1}^m l_{ij}^k=\argmin_{i \in \{2,3\} } \sum_{j=1}^m l_{ij}^k=2$$
\end{itemize}

and transmits it to the hearer. 



%
%

\section{Measures and simulation protocol}

\subsection{Two measures}

To explicitly describe the dynamics under different lexical interests, two measures are defined: the \textit{size of the effective vocabulary} \cite{FS03}

\begin{equation}
V(t)=\frac{1}{np} \sum_{k=1}^p |i: \sum_{j=1}^m l_{ij}^k > 0|
\end{equation}

where $\sum_{j=1}^m l_{ij}^k > 0$ means that the $i$-th word of the lexical matrix $L^k$ is being occupied; and the \textit{energy-like function} $E_{KL}$, defined as

\begin{equation}
E_{KL}(\wp)=d(P(\wp),P(0))+d(P(\wp),P(1))
\end{equation}

where $d$ is the symmetric distance defined by the Kullback-Leibler divergence $KL$ \cite{bishop2006pattern}: $$d(P(\wp),P(0))=KL(P(\wp),P(0))+KL(P(0),P(\wp))$$ Here, $P(\wp)$ denotes the decreasing distribution of frequency meanings for the parameter $\wp$. In order to define the probability distribution $P(\wp)$, a property is imposed to the ranked frequencies $p_i^{\wp}$: $\sum_{i=1}^n p_i^{\wp}=1$, for all $i \in \{1,...,n\}$. 

%

\subsection{Parameters of the simulation}

The analysis is focused on a homogeneuos mixing of agents, the so-called \textit{mean-field} approximation, in which the population is not structured. In this kind of dynamics, at each time step two agents are selected uniformly ar random (one speaker, one hearer). Additionally, as a simple comparison it is described the dynamics on a periodic ring of size $p=128$. For each speaker-hearer interaction, if the speaker is located at the position $k$ the hearer is selected uniformly at random from the \textit{neighborhood} $(k-r,k+r)$, where the radius $r$ varies in $\{1,2,4,8,16,32,64\}$. For all our parameters, the focus here relies on the values after $2p\times 10^4$ speaker-hearer interactions, which average over 10 initial conditions and the last $2 \times 10^3$ steps. One initial condition supposes that each lexical matrix entry is 0 or 1 with probability 0.5. For these measures, the parameter $\wp$ is varied from 0 to 1 with an increment of $3\%$. Each agent is endowed with a $64\times 64$ \textit{lexical matrix}.
%
%
%
\section{Results}


%







Several aspects are remarkable in the behavior of $\langle V \rangle$ versus $\wp$, as shown in Fig. 1. In the first place, the dynamics under the \textit{mean-field} approximation (which is equivalent to select the radius $r$ at random) exhibits three clear domains. First, $\langle V \rangle$ reaches a value close to 1 for $\wp<0.3$. Second, for a critical range of the parameter, $\wp_c \in (0.3,0.6)$, it occurs a transition in which  $\langle V \rangle$ seems to diminish to 0. Finally, for $\wp>0.6$ the dynamics reaches a stationary value $\langle V \rangle \approx 0$.

Qualitatively, the dynamics on one-dimensional rings reproduces the \textit{mean-field} approximation only for small values of $r$. Indeed, for $r<16$ the dynamics seems to qualitatively reproduce the three domains described for the \textit{mean-field} approximation. On the contrary, for $r \geqslant 16$ the behavior of $\langle V \rangle$ tends to exhibit a linear decay.

\begin{figure}[h!]
  \begin{center} 
\includegraphics[width=7 cm, height=6 cm]{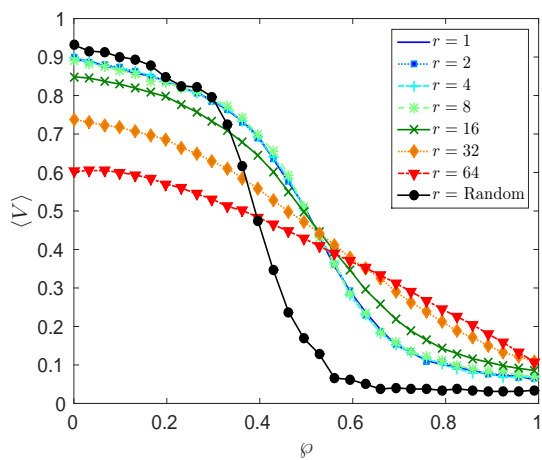}
 
  \end{center}
  \caption{\textbf{$\langle V\rangle$ versus $\wp$}. A population of $p=128$ agents, each one endowed with a $64 \times 64$ lexical matrix, is organized as a one-dimensional ring. Each agent interacts with a neighbor selected (i) at random from the entire ring; or (ii) within a neighborhood defined by a radius  $ r \in \{1,2,4,8,16,32,64\}$. }\label{VPradius}
  \end{figure}
%
%

\begin{figure}[h!]
  \begin{center} 
\includegraphics[width=7 cm, height=6 cm]{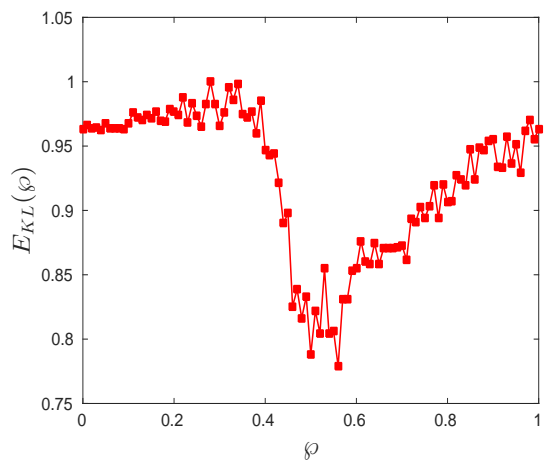}
   \end{center}
  \caption{\textbf{Energy-like function $E_{KL}$ versus $\wp$ is showed}. On a population of $p=128$ agents, each one endowed with a $64 \times 64$ lexical matrix, it is described the behavior of $E_{KL}$ versus $\wp$ for the dynamics under the mean-field approximation.}\label{EvsP}
\end{figure}

One of the most interesting results is summarized by Fig. \ref{EvsP}. At the critical value $\wp_c \approx 0.5$, the energy-like function $E_{KL}$ is minimized for the dynamics under the mean-field approximation.\\ 

%
%

\section{Discussion}

\begin{figure}[h!]
  \begin{center} 
 \includegraphics[width=7 cm, height=6 cm]{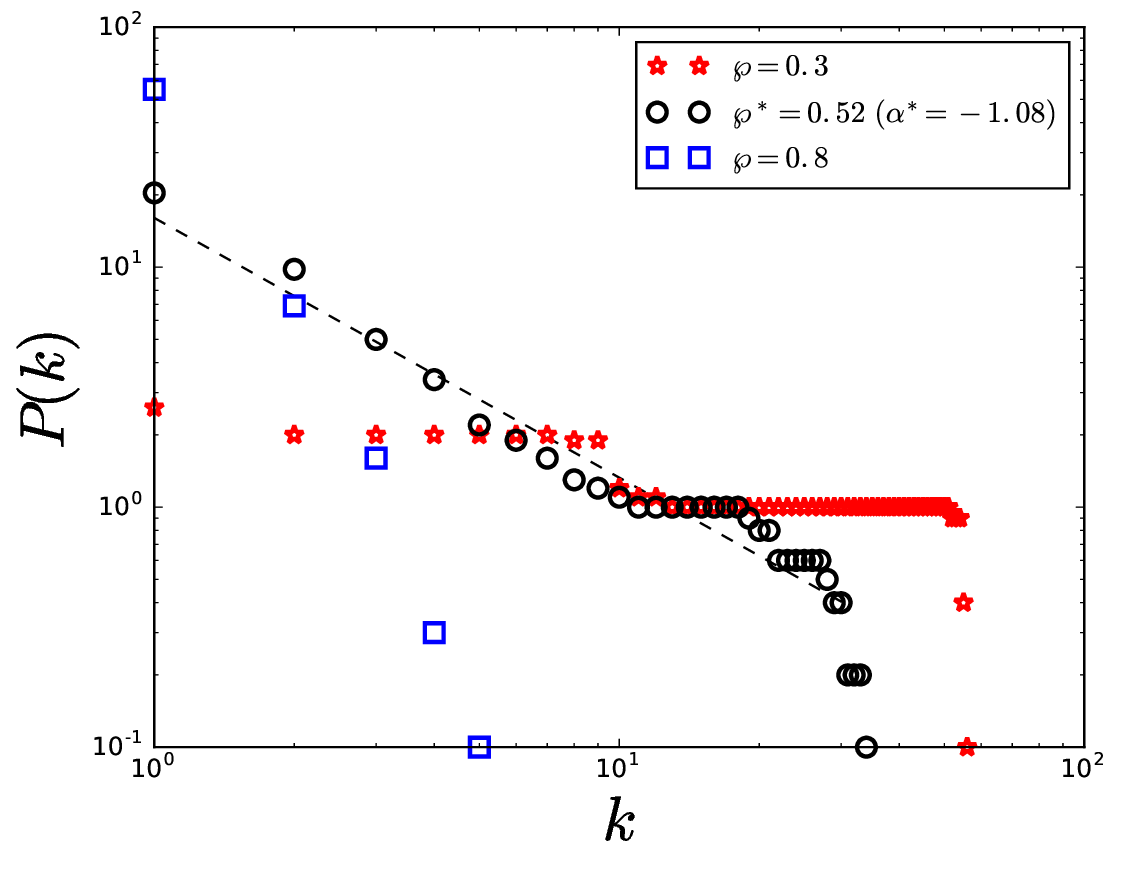}
   \end{center}
  \caption{\textbf{Appearance of an intermediate frequency distribution in vocabularies}. On a population of $p=128$ agents, each one endowed with a $64 \times 64$ lexical matrix, is shown the behavior of $P(k)$ versus $k$ for the mean-field approximation. For $\wp=0.3,0.8$ and the parameter associated to the power law parameter closest to 1 ($\wp_c=0.52$), the figure exhibits the distribution of the number of meanings associated to the $k$-ranked word of the effective vocabulary, $P(k)$, versus $k$ ($\log-\log$ plot). Black depicted lines indicate least squares fit. The calculations average over ten initial conditions. At the critical parameter $\wp_c$, the distribution restricted to the words associated at least to one meaning follows $P(k)\sim k^{-\alpha^{*}=1.08}$. }\label{loglog}
\end{figure}

This work summarizes a decentralized route to the origins of scaling properties in a human-like language. The paper describes particularly the influence of a parameter that measures agents's lexical interests during language game dynamics. The appearance of an intermediate distribution of frequencies at some critical values of the parameter suggests that on a population of artificial agents the emergence of scaling partly arises as a self-organized process only from the pressures of local interactions between agents endowed with intermediate levels of lexical interest (for another view on scaling, see Fig. \ref{loglog}). In some sense, if cooperation is understood as the capacity of selfish agents to forget some of their potential to help one another \cite{Nowak:2006p1455}, the emergence of scaling is crucially influenced by the cooperation between agents. \\
Many extensions of the proposed model should be studied in order to increase the complexity of the language emergence task. A first natural extension should describe more complex ways to define intermediate agent's interests. A second extension should deal with the miniature artificial language learning paradigm to provide direct experimental evidence for the appearance of scaling for real language users optimizing their lexical interests \cite{KANWAL201745}. 

Further research could involve the problem of why the dynamics over one-dimensional rings involve a maximization of the function $E_{KL}$, unlike the dynamics for the mean-field approximation. One preliminary answer is related to the existente of an optimum range of values for $E_{KL}$. 

\section{Acknowledgments}
F.U thanks FONDECYT Chile for financial support under the grant $3180227$. Powered@NLHPC: This research was partially supported by the supercomputing infrastructure of the NLHPC (ECM-02).
 
\bibliographystyle{unsrt}
\bibliography{LaTeXPDFMainDocument}

\end{document}